\documentclass[a4paper,11pt]{article}
\pdfoutput=1 

\usepackage{jcappub} 

\usepackage[T1]{fontenc} 
\usepackage{lineno}

\usepackage{color,xcolor}
\definecolor{officegreen}{rgb}{0.0, 0.5, 0.0}

\title{Prospects for the Observation of Primordial Black Hole evaporation with the Southern Wide Field of View Gamma-ray Observatory}

\author[a,b]{R. L\'opez-Coto}
\author[a,c]{M. Doro}
\author[a,c]{A. de Angelis}
\author[a,c]{M. Mariotti}
\author[d]{J. P. Harding}

\affiliation[a]{Istituto Nazionale di Fisica Nucleare, Sezione di Padova, I-35131, Padova, Italy.}
\affiliation[b]{now at Instituto de Astrof\'isica de Andaluc\'ia, CSIC, 18080 Granada, Spain.}
\affiliation[c]{Universit\`a di Padova, Dipartimento di Fisica, I-35131, Padova, Italy.}
\affiliation[d]{Los Alamos National Laboratory, Los Alamos, USA.}

\emailAdd{rlopez@pd.infn.it}

\abstract{Primordial Black Holes (PBHs) are remnants of objects formed in the early Universe. Their lifetime is an increasing function of their mass, so PBHs in the right mass range can end their lives in an evaporation event that is potentially detectable by our instruments now. This evaporation may result in a $\gamma$-ray flash that can be detected by the current generation of Very-High-Energy $\gamma$-ray detectors. The Southern Wide field of view Gamma-ray Observatory (SWGO) will be part of the next generation of these instruments. It will be able to establish limits on PBH evaporations for integration windows between 0.5 and 5 s, in a radius of 0.25 pc around the Earth, being sensitive to a rate of the order of $\sim$50 pc$^{-3}$ yr$^{-1}$, more than one order of magnitude more constraining than the currently established best limits.}

\begin{document}
\maketitle
\flushbottom
\section{Introduction}
\label{sec:intro}

Black Holes (BHs) radiate particles through their lifetime via the Hawking radiation mechanism \cite{Hawking_BH_evaporation}. This radiative process causes the BH to lose mass over time, with a remaining lifetime that depends on its mass. Primordial Black Holes (PBHs) are BHs whose nature is similar - but whose origin is different - than that of the stellar BHs. These latter appear after the collapse of overdensities of baryonic matter contracted to volumes with radii smaller than the Schwarzschild radius $r_{\rm{s}}=2\,G\,M/c^{2}$ where $G$ is the gravitational constant, $M$ is the BH mass and $c$ the speed of light. PBHs instead are believed to form in the early universe, during a radiation-dominated phase (thus non-baryonic), as the result of quantum mechanical phase transition or collapse of primordial overdensities related to vacuum quantum fluctuations \cite{Hawking_PBH}. The family of theories for the formation of PBHs is wide and the reader is referred to the recent review by~\cite{Sasaki:2018dmp} and references therein.

Differently than stellar BHs, whose seed mass is of the order of the progenitor star mass, the prediction for the seed mass of PBHs spans over several orders of magnitude, due to the fact that their birth mass is related to the time of formation ($t_{\rm H}$) by the relation:

\begin{equation}
\label{eq:pbh_mass}
M_{\rm{PBH}} \sim \frac{c^3\,t_{\rm{H}}}{G} \sim \left( \frac{t_{\rm{H}}}{10^{-23}\;\rm{s}}  \right) \;10^{15}\;\rm{g}
\end{equation}

There is ample literature about two aspects related to the PBH mass: the width of their distribution, and the accretion on this initial seed value. There are many studies that assume different mass distributions~(Eq.~\ref{eq:pbh_mass}) due to the fact that the PBH is formed out of the power spectrum of quantum fluctuations in the early universe, any feature in such spectrum propagates into the PBH mass distribution~\cite{Young_2020}. In addition, PBHs can attract matter as any other BH, and for some PBHs in dense environments, the accretion could have been significant~\cite{Manshanden:2009}, specially for massive PBHs.

Regardless their origin, BHs share some commonalities about their evolution. BHs life expectancy, temperature, mass, and radiative properties are a function of their age.
A PBH with mass $M_{\rm{PBH}}$ (hereafter $M$) has a temperature~\citep{Halzen:1991uw}:

\begin{equation}\label{eq:pbh_temperature}
    T_{\rm{BH}}(M)=\frac{\hbar\,c^3}{8\pi\,G\,k_{\rm{b}}}\frac{1}{M}
    \sim 100 \,\left(\frac{10^{15}\rm{g}}{M}\right) \;[\mbox{MeV}]
\end{equation}
where $\hbar$ is the reduced Planck constant, $k_{\rm{b}}$ the Boltzmann
constant. The temperature increases over time as long as mass is lost, so that the life
expectation for a BH, also called the {\it evaporation time} ($\tau_{\rm BH}$) is:
\begin{equation}
\label{eq:pbh_lifetime}
    \tau_{\rm BH}(M)=\frac{G^2M^3}{\hbar\,c^4}
    \sim 10^{10} \left(\frac{M}{10^{15}\rm{g}}\right)^3 \;[\rm{yr}]
\end{equation}
The reason to use a reference mass of $M_\dagger=10^{15}$g
($\sim10^{-18}$M$_\odot$)  is that Eq.~\ref{eq:pbh_lifetime} shows
that all PBHs roughly lighter than this value would have already
evaporated at present times. Using Eq.~\ref{eq:pbh_lifetime} we can rewrite Eq.~\ref{eq:pbh_temperature} in function of the lifetime as in Ref.~\citep{Ukwatta_2016}: 
\begin{equation}\label{eq:pbh_temp_tau}
    k_b T_{\rm{BH}}(\tau_{\rm{BH}})\approx \left[4.8\times10^{11}\frac{1\,\rm{s}}{\tau_{\rm{BH}}(M)}\right]^{1/3}\;[\mbox{GeV}] 
\end{equation}

If PBHs have an initial mass of around 10$^{15}$ g, they would be evaporating today, producing an increasing emission, culminating with a disruption and a burst of Very-High-Energy (VHE) gamma-ray radiation~\cite{MacGibbon_90, MacGibbon_91}, lasting however short time, from seconds to tens of seconds, but with an extremely high intensity. The expectation for the $\gamma$-ray spectrum is that of the superposition of two components: a primary component coming directly from Hawking radiation, thus peaked at around the PBH mass, and a secondary component, coming from the decay of hadrons produced in the fragmentation of primary quarks and gluons, peaking at somewhat lower energies \cite{Ukwatta_2016}. 

Despite the relatively high gamma-ray flux during evaporation, such disrupting events would not be easy to catch with pointed $\gamma$-ray instruments because of their short duration and random location, preventing external alerts from other observatories to trigger these observations. The search for PBH evaporation has regained interest after the discovery of gravitational waves from the merger of two stellar-mass BHs. It was calculated that BHs compatible with the event could constitute a non-negligible fraction of dark matter \cite{Abbott_2019, Khalouei_2020}. A discovery of PBHs would constitute a major breakthrough, shedding light into the early Universe, cosmology, quantum mechanics mechanisms, particle physics, and BH thermodynamics. Conversely, limits on their number density could translate into relevant boundaries. 

Several studies have found constraints in the number of PBHs within a certain mass range as well as upper limits on the background distribution of PBHs of certain masses using several phenomena as femtolensing~\cite{femtolensing_limits}, microlensing~\cite{Subaru_limits}, capture by Neutron stars~\cite{PBH_limits_2013} and several others as primordial nucleosynthesis, CMB anisotropies, MACHO searches or the search for Hawking radiation (see \cite{carr_limits_pbh_2010} for a summary). There are very constraining limits from PBHs of higher mass from gravitational waves. If these PBHs existed, they should be producing gravitational waves detectable by LIGO/Virgo in their merging. The most constraining lower limits for masses around 10$^{15}$ g come from the measurement of the $\sim$100 MeV Extragalactic Gamma-ray Background (EGB)~\cite{gamma_rays_pbh}. These PBHs should have already evaporated and if their density was high enough, their signature should be present in the EGB which produces a limit on the corresponding cosmological average PBH burst rate density of $<$ 10$^{-6}$ pc$^{-3}$ yr$^{-1}$. On galactic scales, if PBHs are clustered in the Galaxy, we would expect to see an enhancement in the local PBH density and anisotropy in the 100 MeV gamma-ray measurements. Indeed, such an anisotropy has been measured and results in a corresponding Galactic PBH burst limit of $<$ 0.42 pc$^{-3}$ yr$^{-1}$. On the kiloparsec scale, the Galactic antiproton background can be used to give a PBH burst limit of $<$ 1.2 $\times$10$^{-3}$ pc$^{-3}$ yr$^{-1}$. However, the antiproton-derived limit depends on the assumed PBH distribution within the Galaxy and the propagation of antiprotons through the Galaxy, as well as the production and propagation of the secondary antiproton component produced by interactions of cosmic-ray nuclei with the interstellar gas. On parsec scales, the PBH burst limits are directly set by searches for the detection of individual bursting PBHs and are independent of assumptions of PBH clustering~\cite{Ukwatta_2016}. A key aspect is to be able to distinguish a BH burst from GRBs and, in that regard, it is particularly important to remark a soft-to-hard evolution instead of the hard-to-soft one that gamma-ray bursts exhibit \cite{Ukwatta_2016}. 
There are other proposals, such as the identification of closeby short GRBs that would naturally be explained by PBHs \cite{Ukwatta_short_GRBs}.

\subsection{Previous limits on PBH gamma-ray flashes}
In the VHE $\gamma$-ray regime, there have been limits established by several experiments in the past \cite{Tibet_PBH, CYGNUS_PBH, Whipple_PBH, Milagro_PBH}. More recently, there are limits established by Imaging Atmospheric Cherenkov Telescopes (IACTs) currently in operation such as HESS \cite{HESS_PBH, HESS_PBH_2019} and VERITAS \cite{VERITAS_PBH_2017, VERITAS_PBH_2019}. These experiments have established upper limits in PBH evaporation at the level of $>10^4~$pc$^{-3}$ yr$^{-1}$. Besides IACTs, there are also two instruments with a wider Field of View (FoV) working at lower and higher energies than IACTs in the gamma-ray regimen: the $Fermi$-LAT $\gamma$-ray satellite, which recently established a limit on $7 \times10^3$ pc$^{-3}$ yr$^{-1}$ but for much longer time elapse until evaporation (up to 4 years), and a maximum reach $r_{\rm{max}}$=0.02 pc \cite{Fermi_PBH}, and the HAWC air shower array using the water Cherenkov technique to detect gamma rays. HAWC, with the initial 3 years of data, established upper limits at the level of $3.4 \times10^3$ pc$^{-3}$ yr$^{-1}$~\cite{HAWC_PBH}. $Fermi$-LAT and HAWC have a big advantage on the size of the sampled volume thanks to their large FoV and integration time, but they do not reach so far away distances as IACTs.

In this paper, we will study the capabilities to detect PBH flashes of the Southern Wide field of view Gamma-ray Observatory (SWGO), a future air shower array focused in the detection of VHE gamma rays. SWGO is currently in its design phase, and is planned to be composed by a very high fill-factor inner array and a large area low fill-factor component.

\section{Methodology}

\begin{figure}[!h]
\centering
\includegraphics[width=0.95\linewidth]{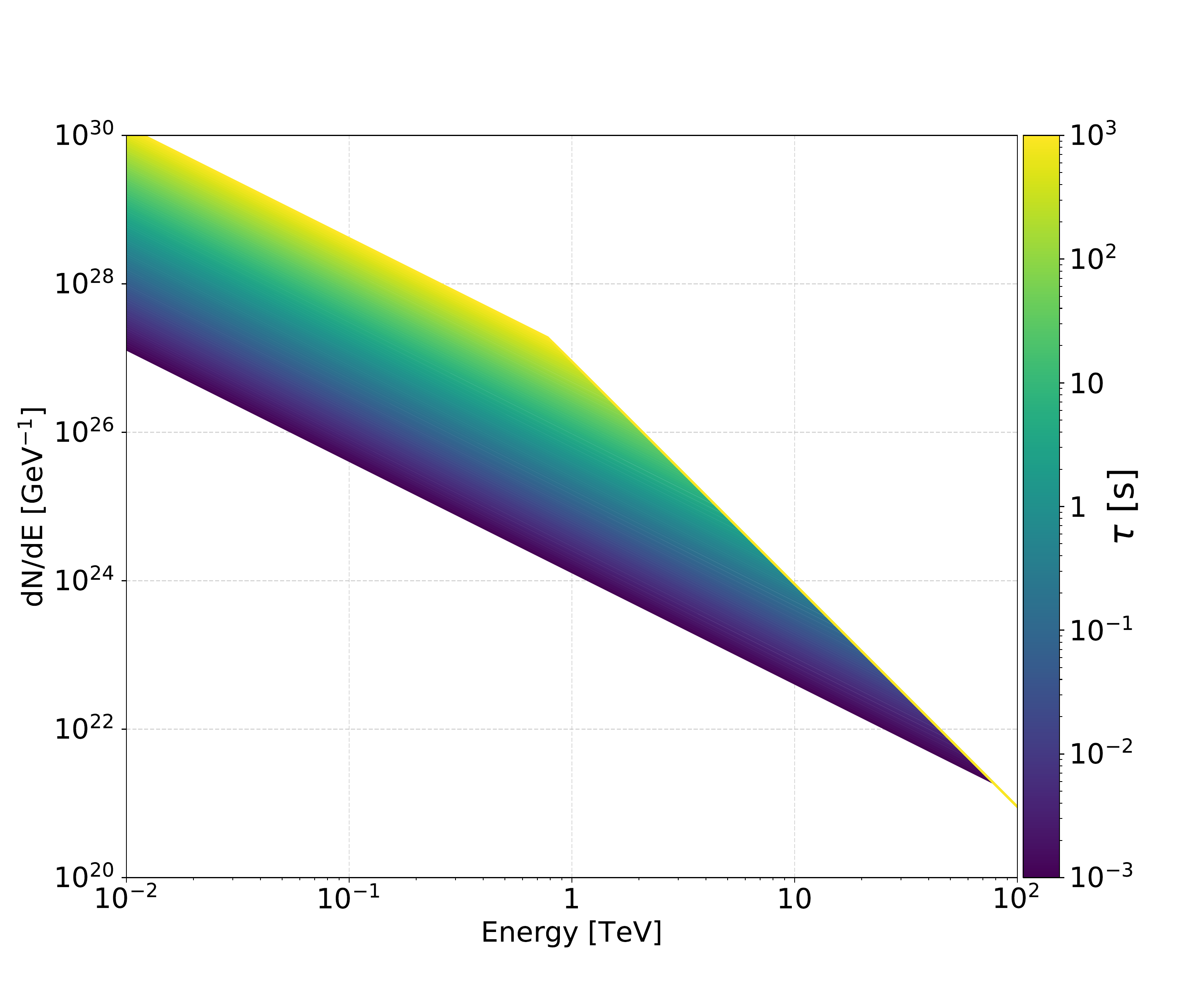}
\caption{Intrinsic spectrum for different burst durations.}
\label{fig:different_duration_burst}
\end{figure}

For the final stages of the life of a PBH we assume the Standard Evaporation Model (SEM) \cite{MacGibbon_90, MacGibbon_91}, in which the emitted photon flux depends on the number of degrees of freedom of the quarks and gluons used in calculating the photon spectrum. The number of photons per unit energy dN$_{\gamma}$/dE, for different burst duration, including the $u$ and $d$ quarks, their antiquarks, and all gluons to calculate the degrees of freedom of the system is given by \citep{2008AstL...34..509P}:

\newcommand\mycom[2]{\genfrac{}{}{0pt}{}{#1}{#2}}
\begin{equation}
\label{dNdE}
    \frac{\rm{d}N_\gamma}{\rm{d}E} \approx 9 \times 10^{35} \begin{cases} \Big(\frac{1~\mathrm{GeV}}{T}\Big)^{\frac{3}{2}} \Big( \frac{1~\mathrm{GeV}}{E}\Big)^{\frac{3}{2}} \makebox[0.4cm]{} \mathrm{GeV^{-1}}\enspace & E < T_{\rm BH} \\
    \enspace\enspace\enspace\enspace\enspace\enspace \Big(\frac{1~\mathrm{GeV}}{E}\Big)^{3} \makebox[0.4cm]{} \mathrm{GeV^{-1}}\enspace & E \geq T_{\rm BH}
    \end{cases}\enspace,
\end{equation}

where $T_{\rm BH}$ is given by Eq. \ref{eq:pbh_temp_tau} and $N_{\gamma}$ is the number of photons. Eq. \ref{dNdE} is represented in Fig. \ref{fig:different_duration_burst} for different burst durations $\tau_{\rm BH}$)(hereafter $\tau$). 

We search for the optimal limits considering different time windows to find the burst duration for which we can establish the most constraining limits in accordance with the performance of our detector. We will discuss the search for signal in the section describing the results achieved.
The number of gamma rays ($\mu$) detected as a function of the distance \textbf{r} and the evaporation time $\tau$ is given by:

\begin{equation}
\label{eq:mu}
\mu(r,\,\tau) = \frac{1-f}{4\pi r^2} \int_{E_1}^{E_2}{ dE'} \int_{0}^{\infty}{{\rm{d}}E \frac{{\rm{d}}N(\tau)}{{\rm{d}}E} A(E) G(E,E')}
\end{equation}
where $f$ is the dead time of the detector (ignored in this study, and only relevant for short duration searches), $E'$ is the estimated energy, $E$ is the true energy, $E_1$=10 GeV, $E_2$=100 TeV are the integration limits in estimated energy, $A$ the effective area of the detector and $G$ the function accounting for the detector resolution.

The pre-trial probability ($p_{\rm pre-trial}$) that is needed to obtain a 5$\sigma$ post-trial probability of detection over the average background ($p_{\rm post-trial} = 2.89\times10^{-7}$ $(5\sigma)$), considering trials $N_{\rm t}$ can be written as:
\begin{equation}
\label{eq:post-trial-complete}
p_{\rm{post-trial}} = 1-(1-p_{\rm pre-trial})^{N_{\rm t}}
\end{equation}

and approximated to:

\begin{equation}
\label{eq:post-trial}
p_{\rm{pre-trial}} \approx \frac{p_{\rm post-trial}}{N_{\rm t}}
\end{equation}

where the number of trials for SWGO depends on the search window duration $T_{\rm search}$:
\begin{equation}
N_{\rm t} (\tau) = \frac{T_{\rm search}}{\tau} \left(\frac{\theta_{\rm{fov}}}{\theta_{\rm{res}}} \right)^2
\end{equation}
where $\tau$ the total duration of the burst and $\theta_{\rm{fov, res}}$ are the total FoV and the angular resolution of SWGO, respectively. Following Eq. \ref{eq:mu} and considering a sphere of radius $r$, the maximum distance at which a source can be detected would be:

\begin{equation}
\label{eq:rmax}
r_{\rm{max}} = \sqrt{\frac{1-f}{4\pi\mu(\tau)}\int_{E_1}^{E_2}{ dE'} \int_{0}^{\infty}{dE \frac{dN(\tau)}{dE} A(E) G(E,E')}}
\end{equation}
and the volume probed by the detector is then:

\begin{equation}
V(\tau) = \frac{4\pi r_{\text{max}}^3(\tau)}{3} \frac{\text{FoV}_{\rm SWGO}}{4\pi} 
\end{equation}

where FoV$_{\rm SWGO}$ is the total solid angle covered by SWGO. The upper limit at a 99\% Confidence Level on the number of PBHs evaporating per unit volume is finally:

\begin{equation}
\text{UL}_{99} = \frac{4.6}{V \cdot T_{\rm search}}
\end{equation}

where the 4.6 is the upper limit of observing 0 bursts with a 99 \% confidence level: P(0|4.6) = 0.01~\cite{Milagro_PBH}.

For the Instrument Response Functions (IRFs) of SWGO, we use the ``straw-man'' design presented by \cite{SGSO_white_paper}: a detector located at 5,000 m altitude composed of a compact inner detector covering 80,000 m$^2$ with an 80\% coverage and a sparse outer detector covering 221,000 m$^2$ with 8\% coverage. The ``straw-man'' design is detector-agnostic, and the IRFs have been calculated using atmospheric shower simulations counting the number of particles and energy deposited on the ground. The effective area, angular resolution and energy bias and resolution considered here are the ones presented in Figures 3.1 and 3.2 of \cite{SGSO_white_paper}. These IRFs were derived using the available simulations at 20 deg zenith, with a $\gamma$-ray efficiency of 75\% using the the sensitivity tools developed for \cite{DM_SWGO} \footnote{https://github.com/harmscho/SGSOSensitivity}.

\section{Results}
Let us study the case example of SWGO for a search in the $\tau$ range between 0.1 - 10$^3$ s and three different scenarios: 1, 5 and 10 years of total observation time. We will ignore in this study the dead time effects and overlap between time and angular windows, since they affect the results to a lesser extent than other quantities such as the rate of the detector. We will call here ``background rate'' to the rate of random events that trigger the detector and pass the $\gamma$-ray selection cuts. We consider three cases  for the background rate: the {\it fiducial} case simulated and mentioned above, a {\it conservative} case in which a 50 \% larger background and a {\it optimistic} case in which a  20 \% better PSF is assumed. These different cases allow to bracket the confidence level in the PBH limits derived. We will also consider the angular resolution as a function of the energy to count the number of trials of the detector. To combine the results from the inner and the outer SWGO detector, we use a combined likelihood to obtain the number of gamma rays from Equation \ref{eq:mu}.

\begin{figure}[!h]
\centering
\includegraphics[width=0.95\linewidth]{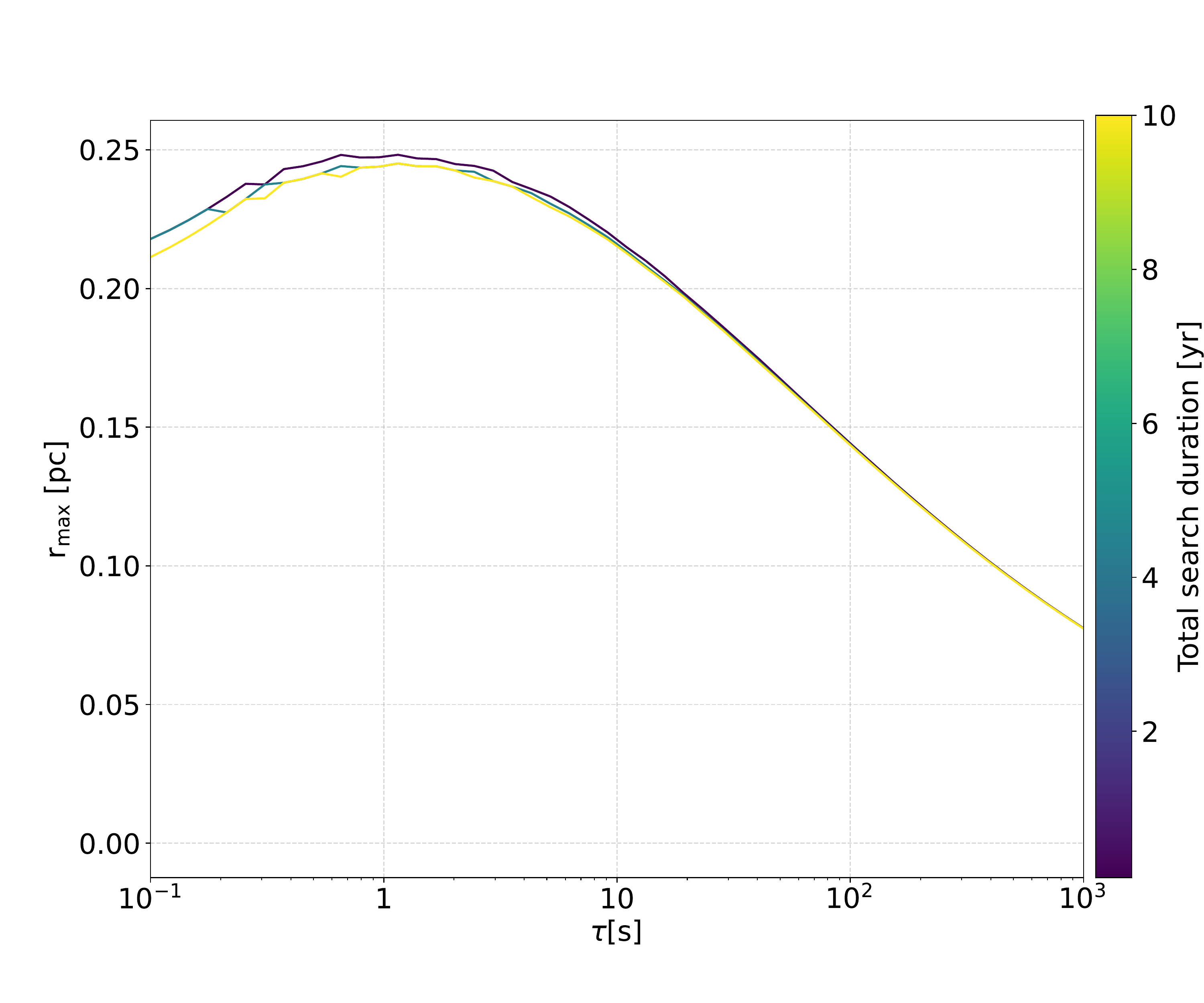}
\caption{$r_{\rm{max}}$ as a function of the burst duration $\tau$, for different total search durations (inner array).}
\label{fig:rmax}
\end{figure}

The $r_{\rm{max}}$ as a function of $\tau$ for different total search durations is shown in Fig. \ref{fig:rmax}. We note that there is not a big difference in the $r_{\rm{max}}$ reached using different total search durations. This is because Eq. \ref{eq:rmax} is only modified by the duration of the search in the number of signal events $\mu$ (Eq. \ref{eq:mu}) to get a 5$\sigma$ post-trial detection. Even though the post-trial probability (Eq. \ref{eq:post-trial}) varies one order of magnitude comparing the total search duration of 1 and 10 years, the number of signal events $\mu$ does not have such a large variation.


Assuming no serendipitous detection, we derive upper limits for different burst durations and different search durations for the cases of 1, 5 and 10 years of observations and draw bands between the conservative and optimistic scenarios in the burst rate limits. It is worth mentioning that the "Burst duration" is a variable that helps us optimize the integration window for our searches and the only relevant quantity is the best limit for a given search duration. 

\begin{figure}[!h]
\centering
\includegraphics[width=0.95\linewidth]{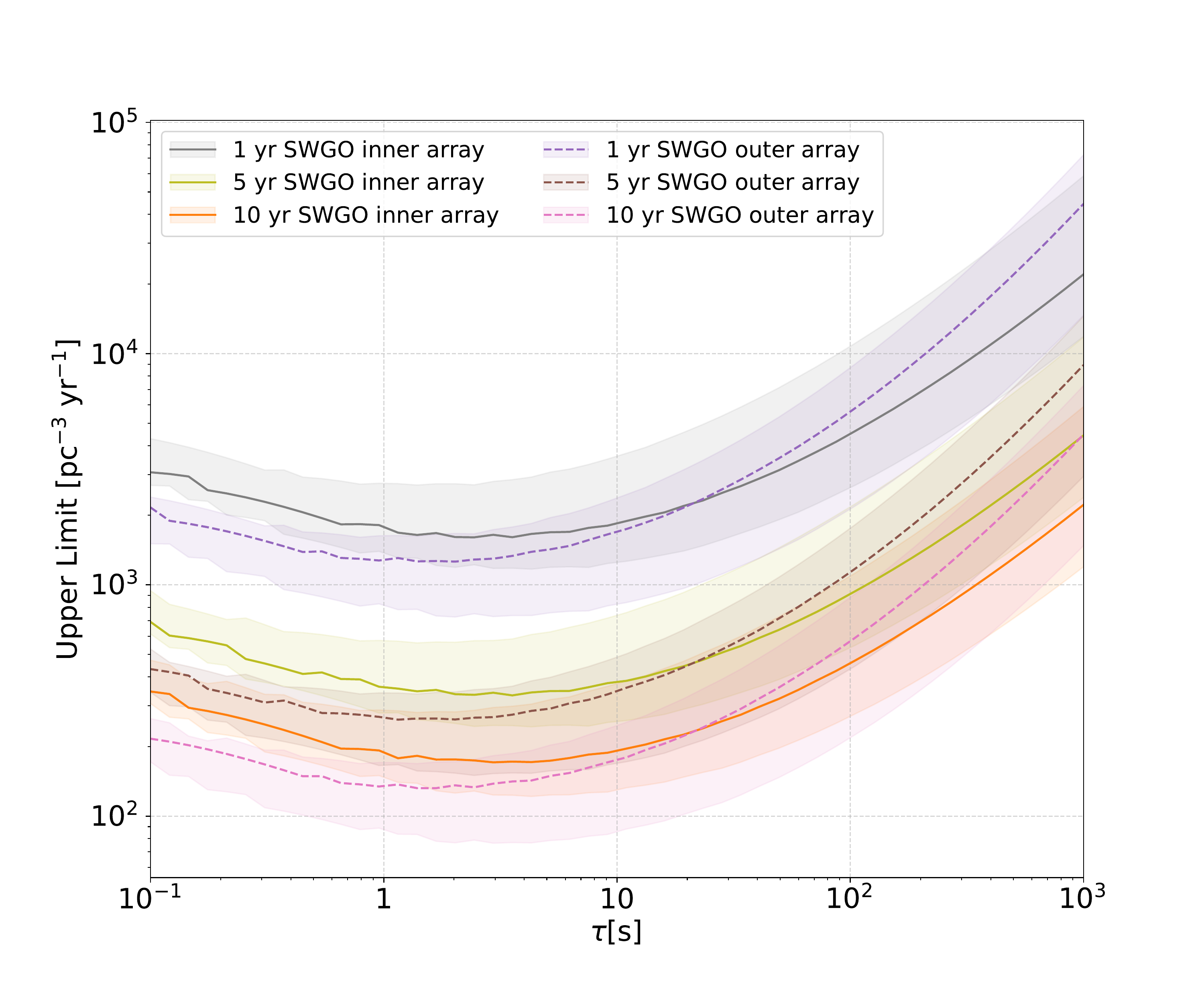}
\caption{PBH limits as a function of the burst search duration $\tau$, for different search durations, for the inner and outer SWGO array.}
\label{fig:pbh_limits}
\end{figure}

It is important to note that, according to Fig. \ref{fig:pbh_limits}, although the most constraining limits are reached for both arrays for search durations between 1 - 10 s, the outer array, less sensitive at the lowest energies, overcomes the performance of the inner array for short durations. The reason is that according to Eq. \ref{dNdE}, for short durations the spectrum is harder and the larger collection area of the outer array at high energies compensates for the low sensitivity at low energies. This implies that a more sparse array with better sensitivity for multi-TeV energies is helpful to increase the sensitivity of PBH searches.

\begin{figure}[!h]
\centering
\includegraphics[width=0.95\linewidth]{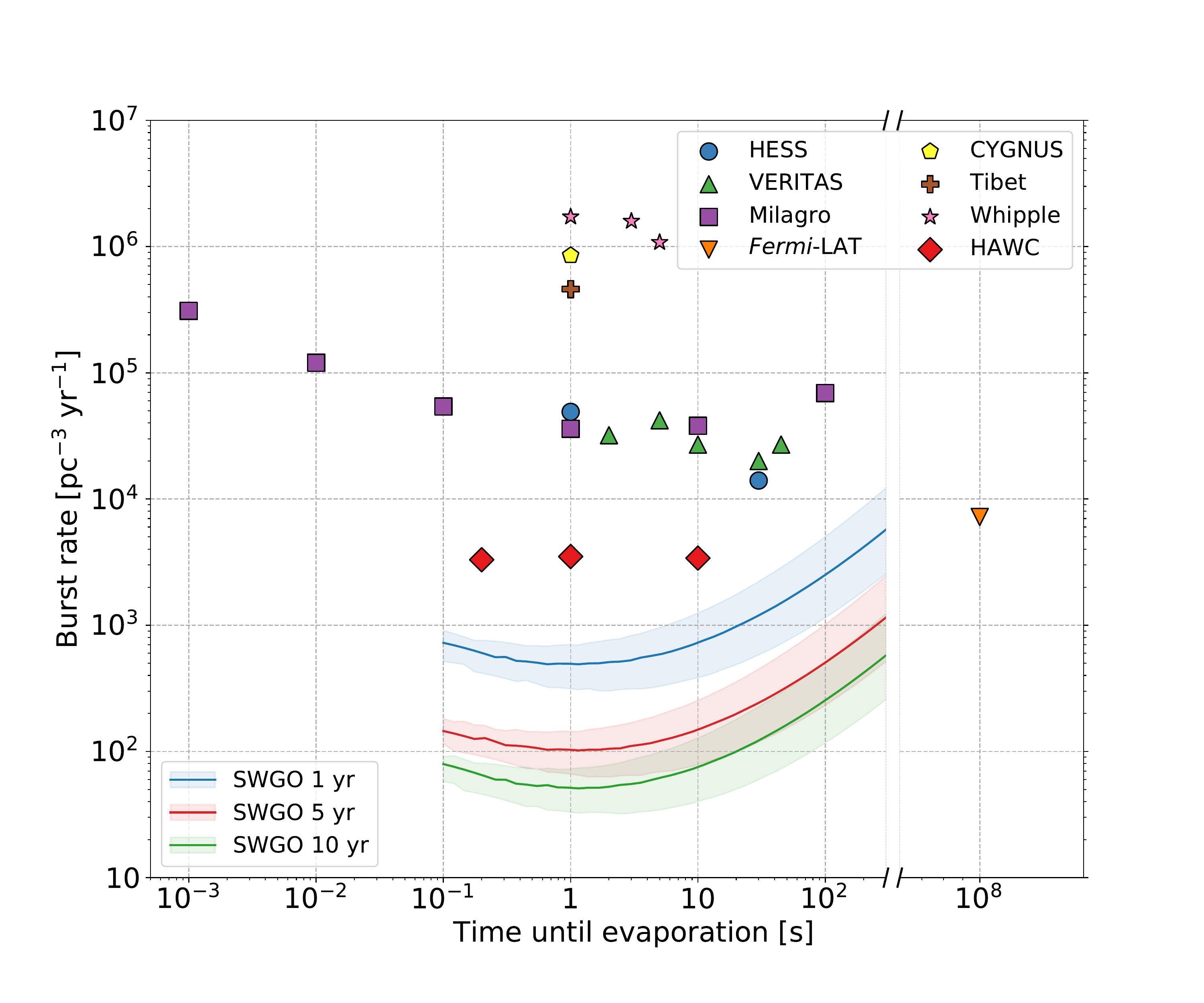}
\caption{SWGO combined sensitivity to PBH bursts of different durations compared to results of different experiments. Data taken from \cite{Tibet_PBH, CYGNUS_PBH, Whipple_PBH, Milagro_PBH, HESS_PBH_2019, VERITAS_PBH_2019, Fermi_PBH, HAWC_PBH}. }
\label{fig:pbh_uls}
\end{figure}

Using the inner and outer array results, we perform a joint likelihood to compute the combined sensitivity to the PBH burst rate. The final comparison of PBH burst limits expected for SWGO with the observations performed by other VHE $\gamma$-ray observatories is shown in Fig. \ref{fig:pbh_uls}.

\section{Discussion}

The best PBH burst limits are reached for search durations from 0.5 to 5 s. For example, for 1-second integration windows in the 10-year search, we obtain $N_t\sim10^{10}$, a number of background events in each angular and time window of $\sim$7 and the needed number of excess events to obtain a post-trial probability of 5$\sigma$ is $\sim$40. We can see that using combined IRFs between the inner and the outer array, we can reach limits on $\gamma$-ray flash duration about one order of magnitude lower than the best ones to date established by HAWC \cite{HAWC_PBH}. 10 years of data should provide limits about two orders of magnitude lower than the reach of the best experiments currently in operation. 

\begin{figure}[!h]
\centering
\includegraphics[width=0.95\linewidth]{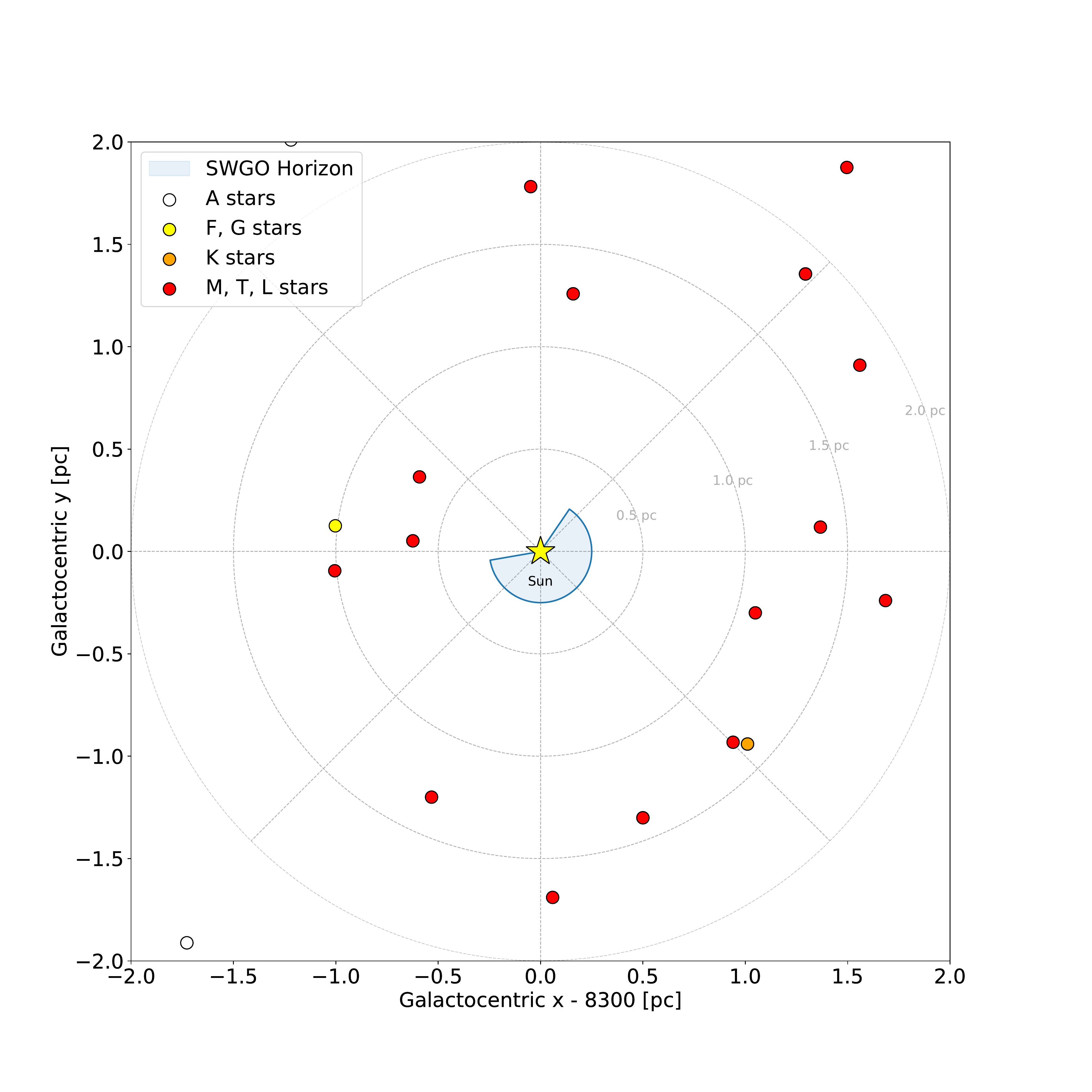}
\caption{Galactocentric horizon sensitivity of SWGO. Also included stars located nearby the Earth.}
\label{fig:galactocentric_swgo_horizon}
\end{figure}

We would like to mention that future experiments and observatories will also be able to overcome the current limits established by HAWC. With a larger  $r_{\rm{max}}$ reach thanks to its better sensitivity in a larger energy range, the Cherenkov Telescope Array will establish better limits than the current generation of IACTs, but still not competitive with wide FoV experiments. The LHAASO experiment, using also a wide FoV technique, has good chances of establishing limits competitive with those foreseen in this paper. LHAASO will however cover the Northern hemisphere, while the region depicted in Fig. \ref{fig:galactocentric_swgo_horizon} will only be covered by SWGO with a burst limit smaller than 50 pc$^{-3}$yr${-1}$.

\section*{Acknowledgements}
This work was financially supported by the European Union's Horizon 2020 research and innovation program under the Marie Sk\l{}odowska-Curie grant agreement No. 754496 - FELLINI. R.L.-C. acknowledges the financial support from the State Agency for Research of the Spanish MCIU through the ‘Center of Excellence Severo Ochoa’ award to the Instituto de Astrofísica de Andalucía (SEV-2017-0709). MD acknowledges funding from Italian Ministry of Education, University and Research (MIUR) through the ``Dipartimenti di eccellenza'' project Science of the Universe. PH acknowledges the support of the US Department of Energy Office of High-Energy Physics and the Laboratory Directed Research and Development (LDRD) program of Los Alamos National Laboratory. The authors would like to thank M. Peloso and F. D'Eramo for useful discussions. The authors would also like to thank the SWGO collaboration for useful discussions during the writing of this paper.

\bibliographystyle{plain} 

\bibliography{./references}

\end{document}